\documentclass[aps,pre,twocolumn,superscriptaddress,amsmath]{revtex4}

\usepackage{graphicx}

\allowdisplaybreaks

\begin{document}

\title{Shape Selection in Chiral Self-Assembly}

\author{Robin L. B. Selinger}
\affiliation{Center for Bio/Molecular Science and Engineering, Naval Research
Laboratory, Code 6900, 4555 Overlook Avenue, SW, Washington, DC 20375}
\affiliation{Physics Department, Catholic University of America, Washington, DC
20064}
\author{Jonathan V. Selinger}
\author{Anthony P. Malanoski}
\author{Joel M. Schnur}
\affiliation{Center for Bio/Molecular Science and Engineering, Naval Research
Laboratory, Code 6900, 4555 Overlook Avenue, SW, Washington, DC 20375}

\date{July 1, 2004}

\begin{abstract}
Many biological and synthetic materials self-assemble into helical or twisted
aggregates. The shape is determined by a complex interplay between elastic
forces and the orientation and chirality of the constituent molecules. We study
this interplay through Monte Carlo simulations, with an accelerated algorithm
motivated by the growth of an aggregate out of solution. The simulations show
that the curvature changes smoothly from cylindrical to saddle-like as the
elastic moduli are varied. Remarkably, aggregates of either handedness form from
molecules of a single handedness, depending on the molecular orientation.
\end{abstract}


\maketitle

A wide range of materials self-assemble into chiral aggregates. These include
biological materials and their synthetic analogues, such as
carbohydrate-~\cite{pfannemuller85} and amino-acid-based
amphiphiles~\cite{nakashima84}, peptides~\cite{zhang93,aggeli97}, diacetylenic
lipids~\cite{yager84,schnur93}, gemini surfactants~\cite{oda99}, and
multicomponent mixtures in bile~\cite{chung93,zastavker99}. Aggregates form as
bilayer membranes in solution, and grow into a range of chiral shapes, including
tubules with ``barber-pole'' markings, helical ribbons with cylindrical
curvature, and twisted ribbons with Gaussian saddle-like
curvature~\cite{spector03}.  They form through a variety of pathways, including
reorganization of large spherical vesicles and growth directly out of
solution~\cite{spector97,thomas99}. In the latter process, aggregates twist in a
helix while they increase in length.

While one might expect that macroscopic handedness of an aggregate is tied to
the molecular-scale chirality of its constituents, experiments have shown that
the relationship is not always that simple.  In the early stages of
self-assembly, diacetylenic lipids can form both right- and left-handed helical
ribbons, even though the equilibrium state has tubules of a uniform
handedness~\cite{thomas99}. In a biological analogy, chirality reversal at the
organism level arises occasionally via genetic mutations, and has been observed
in plants, molluscs, birds, and mammals, even though the proteins and other
molecules within each organism retain their normal
chirality~\cite{souliemarsche99,ueshima03,levin98}.

Experiments have also shown that the relationship between material properties
and the macroscopic shape of chiral aggregates is surprisingly complex.  For
example, charged gemini surfactants with chiral counterions exhibit a transition
between helical ribbons with cylindrical curvature and twisted ribbons with
Gaussian curvature as a function of molecular chain length~\cite{oda99}.
Similarly, mixed bilayers of saturated and diacetylenic phospholipids show
transitions from micron-scale cylindrical tubules to twisted ribbons, and then
to nanometer-scale tubules, as a function of temperature~\cite{spector01}.

To understand the shape of chiral aggregates, and to design new structures for
nanotechnology, we must learn how to control aggregate size and shape by
adjusting the chemical composition and the conditions under which self-assembly
occurs. An important step toward this goal is to understand how the membrane's
elastic moduli, chirality, and tilt orientation together control the structure
of the resulting aggregate. Earlier theoretical models have studied this
by using continuum elastic theory to calculate the energy of twisted membranes
in simple geometries~\cite{selinger01}. Such models show that chiral
interactions, coupled with molecular tilt, lead to the formation of cylindrical
tubules or helical or twisted ribbons. The limitation of this analytical
approach is that it considers only idealized structures, and does not explore
the full range of potential shapes or the mechanism of shape transitions.
Finite-element modeling of membranes and vesicles would allow free exploration
of shapes, but no such models have taken the membrane's molecular tilt and
chiral interactions explicitly into account. Molecular-scale simulation of
tilted bilayer membranes is another possible approach~\cite{hofsaess03}, but
available system sizes are too small to allow studies of shape selection.

In this paper, we present a new mesoscale simulation approach to explore the
roles of elasticity, chirality, and molecular tilt in chiral self-assembly. We
develop a simulation technique motivated by the growth of an aggregate out of
solution, which allows the aggregate to select its shape through a Monte Carlo
process. This approach provides an explicit visualization of a range of possible
shapes, and shows that the aggregate shape varies smoothly from cylindrical to
Gaussian curvature as the elastic properties of the membrane are varied.
Moreover, it shows that the handedness of the aggregate depends on the
orientation of molecular tilt as well as on the handedness of the chiral
interactions, demonstrating one mechanism by which both right- and left-handed
structures can form from a material of uniform molecular chirality. Thus, our
simulations may explain a wide range of experimental findings: the formation of
helical and twisted ribbons, the transitions between these shapes in gemini
surfactants~\cite{oda99} and lipid mixtures~\cite{spector03}, and the
observation of chiral aggregates with both helical senses in the same
lipid~\cite{thomas99}.

\begin{figure}
\includegraphics[clip,width=2.4in]{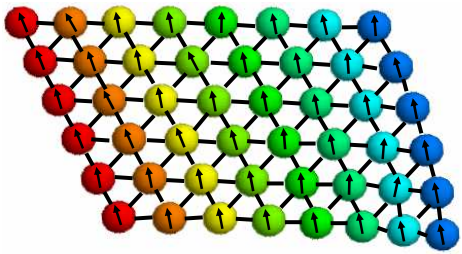}%
\caption{(Color online) Model of a chiral elastic ribbon as a tethered membrane
composed of interaction sites connected by elastic springs to form a triangular
lattice. The local molecular orientation at each site is described by a
director, as indicated by the arrows.}
\end{figure}

In our mesoscale approach, a chiral elastic ribbon is modeled as a tethered
membrane composed of interaction sites connected by elastic springs to form a
triangular lattice, as shown in Fig.~1. Each interaction site represents a small
``patch'' of membrane. The membrane can deform in three-dimensional (3D) space,
so that each site has position $\mathbf{r}_i$, and local upward normal
$\mathbf{n}_i$.  The local molecular orientation at each site is described by
the director $\mathbf{d}_i$, a vector of unit length. This approach is analogous
to earlier studies of tethered membranes~\cite{kantor86}, but here the membrane
has the additional features of chirality and molecular orientation. It is also
analogous to mesoscale models of smectic liquid crystals, which have chirality
as well as a molecular orientation defined on a lattice~\cite{selinger00}, but
here the membrane is free to change its shape in 3D. In our initial exploration
of the model, we constrain the director on each site at a typical experimental
tilt angle of $30^\circ$ with respect to the local upward normal $\mathbf{n}_i$,
with the tilt oriented along the ribbon's long edge, which is also along a bond
between neighboring sites. In future work we will remove this constraint and
consider the possibility of modulated membrane tilt.

To investigate shape selection, we must define a model for the elastic energy of
a chiral membrane that takes into account the microscopic chirality of its
constituent material. We use a lattice expression analogous to the continuum
energy for chiral membranes investigated in earlier work~\cite{selinger01}. It
includes a chiral elastic term as well as bend, stretch, and hard-core
contributions:
\begin{eqnarray}
E&=&\sum_{\langle i,j\rangle} \biggl[
\lambda (\mathbf{d}_i\times\mathbf{d}_j)\cdot
\frac{(\mathbf{r}_i-\mathbf{r}_j)}{|\mathbf{r}_i-\mathbf{r}_j|}
-K_\text{bend}(\mathbf{n}_i\cdot\mathbf{n}_j)\nonumber\\
&&\phantom{\sum_{\langle i,j\rangle} \biggl[}
+\frac{1}{2}K_\text{stretch}\left(|\mathbf{r}_i-\mathbf{r}_j|-R_0\right)^2
\biggr]\\
&&+\sum_i \sum_j U_\text{hc}\left(|\mathbf{r}_i-\mathbf{r}_j|\right).\nonumber
\end{eqnarray}
In the first term of this expression, the chiral elastic parameter $\lambda$
gives an energetic preference for twisting the directors $\mathbf{d}_i$ and
$\mathbf{d}_j$ away from parallel alignment. Reversing the sign of $\lambda$
reverses the favored chiral twist of the director field. In the second term, the
parameter $K_\text{bend}$ controls the energy of bending the membrane so that
the local normal vectors on sites $i$ and $j$ are not parallel. In the third
term, the Hooke's law constant $K_\text{stretch}$ gives the energy of separating
sites $i$ and $j$ away from their equilibrium separation $R_0$. In those three
terms, the sum on $\langle i,j\rangle$ refers to all nearest-neighbor pairs of
sites in the membrane. The final term in the energy represents a hard-core
potential between any pair of sites, with $U_\text{hc}(r)=0$ for $r>a$, and
$U_\text{hc}(r)=\infty$ for $r\le a$, where $a=\frac{2}{3}R_0$ is the hard-core
diameter. That term prevents the membrane from intersecting itself.

For Monte Carlo simulations of the membrane, we need a simulation algorithm that
will come to equilibrium in a reasonable time. For a first attempt, we
initialized the system as a long, flat ribbon, and allowed it to evolve via
off-lattice single-particle moves with the Metropolis algorithm. However, we
found that this process evolves much too slowly to reach a twisted shape.
Because the twist must start at the ribbon's narrow ends and diffuse inward
toward the middle, there is a barrier against twisting a ribbon, and the
equilibration time grows sharply with the ribbon length.

\begin{figure}
\includegraphics[clip,width=2.4in]{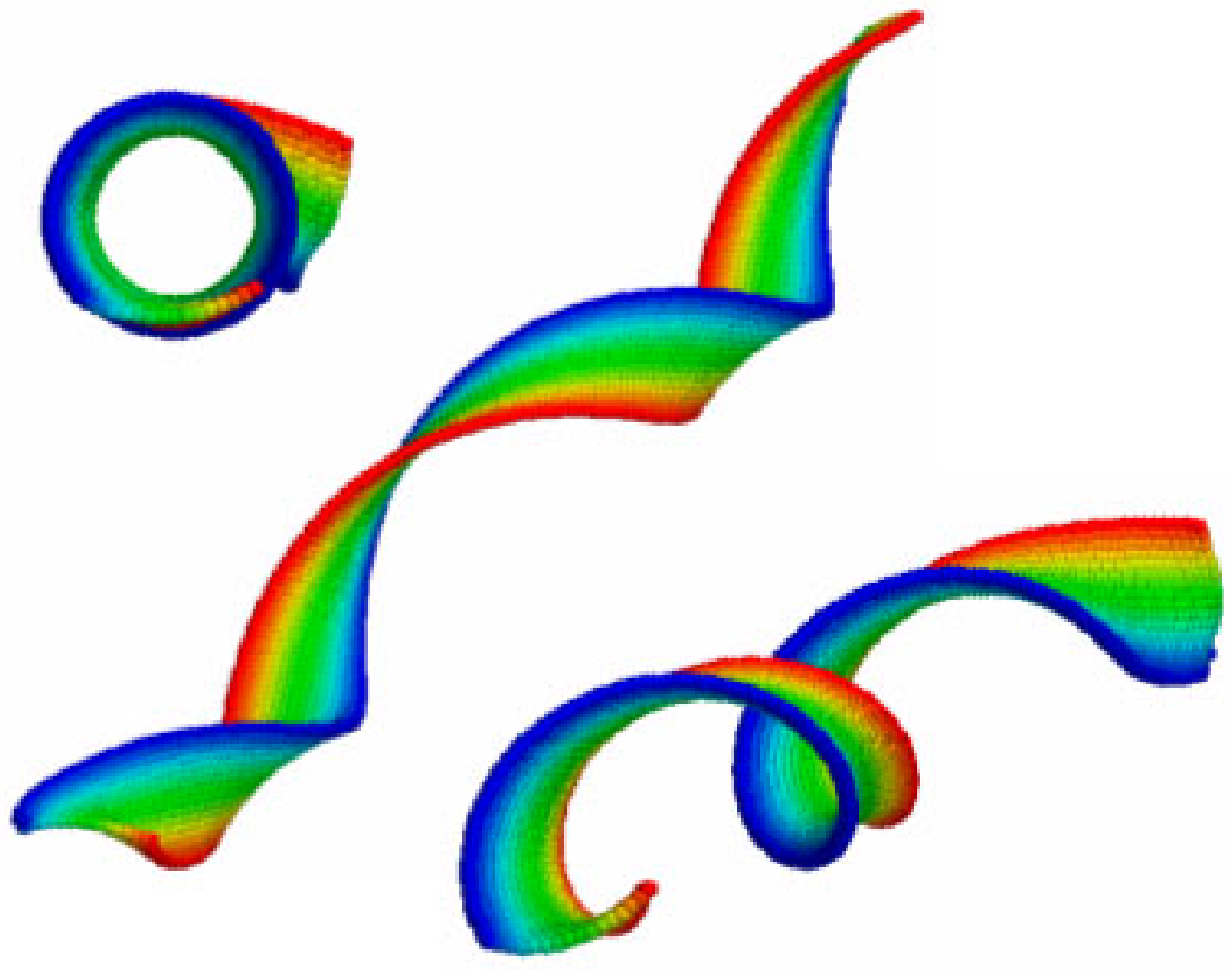}%
\caption{(Color online) Three views of a helical ribbon, showing the 3D
structure. The ribbon has grown to a size of $16\times216$ sites.}
\end{figure}

To develop an alternative simulation approach, we are motivated by one of the
experimental pathways for the formation of chiral aggregates. When ribbons grow
out of solution, they wrap into a helix as they grow, and hence they have no
barrier against twisting. For that reason, we simulate a growth process, with
particles attaching onto one edge of the model aggregate. The concept of
simulating a growth process has found wide application in the study of
polymers~\cite{rosenbluth55,frenkel96}, but this is its first application to
chiral membranes. While this simulation approach mimics an observed growth
mechanism, it is essentially a mesoscopic Monte Carlo algorithm, which does not
take into account any of the detailed kinetics associated with molecular
aggregation.

We start the system as a small ribbon of size $16\times16$ sites, and then add
new rows one at a time up to a maximum size of $16\times216$ sites. After each
row addition, the last 8 rows are equilibrated for 10,000 Monte Carlo steps per
site, and then the whole ribbon is equilibrated for an additional 5000 Monte
Carlo steps per site. The temperature is kept near zero so that the Monte Carlo
process is essentially an energy minimization, which allows comparison with
previous analytical results. In future work we will address the role of thermal
fluctuations in membrane shape. An example of a chiral membrane grown in this
fashion is shown in Fig.~2, and animations of the growth are presented in
EPAPS~\cite{epaps}. In the figures and animations, colors are a guide to the eye
to visualize the 3D structure.

\begin{figure}
\includegraphics[clip,width=2.4in]{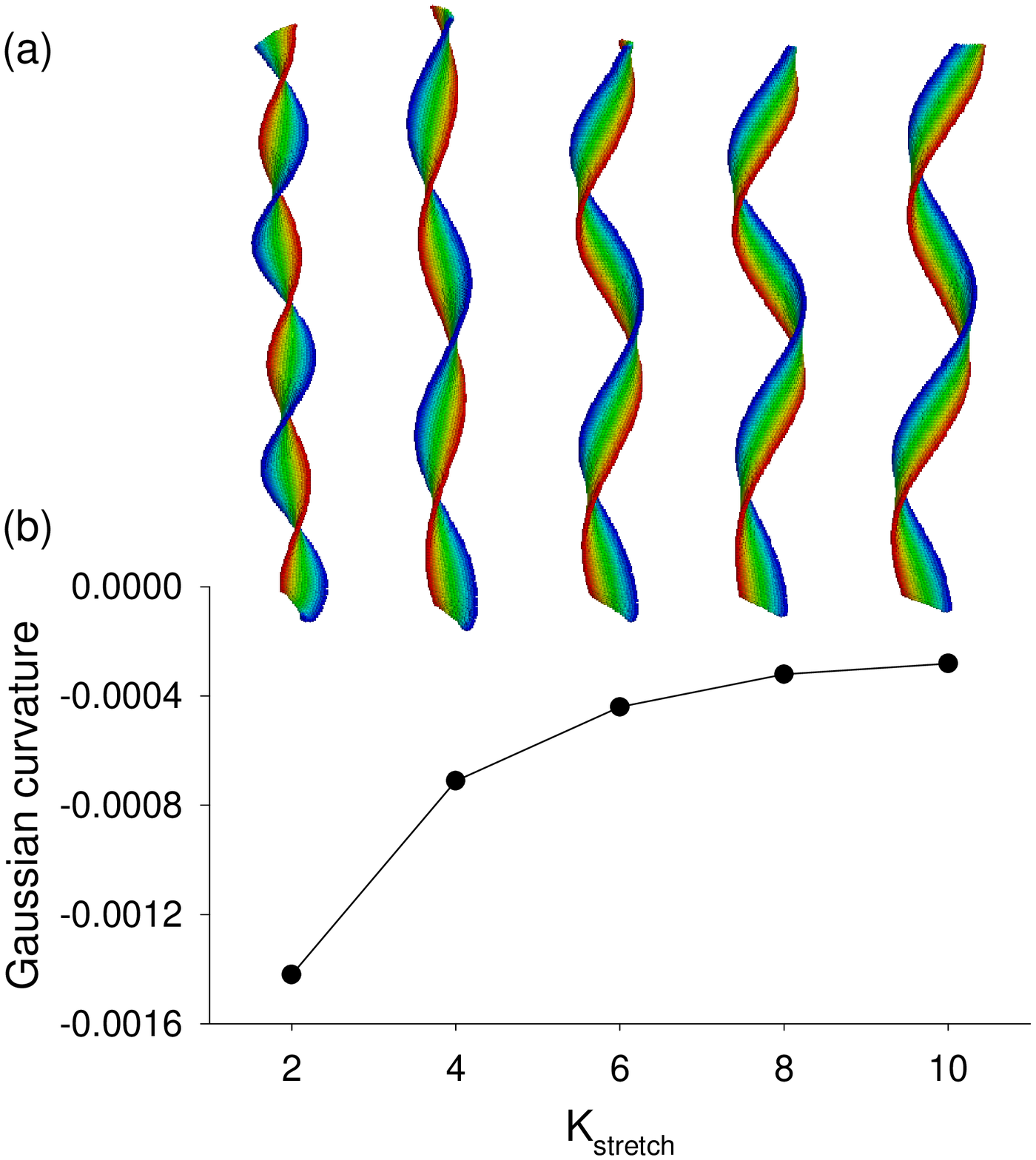}%
\caption{(Color online) (a)~Series of ribbons grown with the parameter
$K_\text{stretch}=2$, 4, 6, 8, and 10, showing the crossover from twisted
ribbons with Gaussian saddle-like curvature to helical ribbons with cylindrical
curvature. (b)~Plot of the Gaussian curvature for the five ribbons, calculated
at the central point of the ribbons, in units of the hard-core diameter
$a^{-2}$. This plot confirms that all of the structures have some Gaussian
curvature, and that the level of Gaussian curvature changes gradually as a
function of the elastic parameters.}
\end{figure}

We use this simulation approach to investigate how the final equilibrium shape
of the aggregate depends on membrane properties. Of the three elastic parameters
$K_\text{stretch}$, $K_\text{bend}$, and $\lambda$, it is necessary to vary two
since only their ratios are significant at low temperature. We begin by
examining the effect of varying the stretch modulus $K_\text{stretch}$ over a
range from 2 to 10 in arbitrary units, while $K_\text{bend}=1$ and $\lambda=1$
are held fixed. We observe the gradual structural transition shown in Fig.~3(a).
For low $K_\text{stretch}$, we find a twisted ribbon with Gaussian saddle-like
curvature, while for high $K_\text{stretch}$, the shape is a helical ribbon with
cylindrical curvature. This transition is physically reasonable, because
Gaussian curvature requires stretching the membrane, which is suppressed by
$K_\text{stretch}$. The small Gaussian curvature that remains in helical ribbons
for high $K_\text{stretch}$ resembles the ripple structure that has been
predicted theoretically~\cite{selinger96}.

We have done an analogous study of the effects of varying the bend modulus
$K_\text{bend}$. We fix $K_\text{stretch}=6$ and $\lambda=1$, and vary
$K_\text{bend}$ over the range from 1 to 2.4. The simulation results (not shown)
give a similar gradual transition from helical ribbons at low $K_\text{bend}$ to
twisted ribbons at high $K_\text{bend}$.

Each structure is characterized by a radius, a pitch, and two curvature
parameters. Through the structural transition shown in Fig.~3(a), all of these
measures show a continuous and gradual transition. In particular, note that none
of the structures are ideal twisted or helical ribbons with pure Gaussian or
cylindrical curvature. Rather, each structure involves a combination of Gaussian
and cylindrical curvature, and the combination changes gradually as a function
of $K_\text{stretch}$. We can calculate the Gaussian curvature at any
interaction site by adding up the six incident bond angles $\theta_i$ at that
site. If the Gaussian curvature is zero, then the six bond angles add up to
$2\pi$. Hence, the deviation of the sum away from $2\pi$ is a simple measure of
the Gaussian curvature~\cite{regge61,lin82}:
\begin{equation}
\kappa_\text{Gauss}=\frac{3\left(2\pi-\displaystyle\sum_{i=1,6}\theta_i\right)}%
{\displaystyle\sum_{i=1,6}a_i},
\end{equation}
where $a_i$ are the areas of the triangles adjacent to the site. Figure~3(b)
plots the Gaussian curvature of each ribbon shown in Fig.~3(a), evaluated at the
central point of the ribbon, as a function of $K_\text{stretch}$. The variation
in Gaussian curvature is very substantial, ranging over a factor of 5. However,
it does not occur as a sharp structural transition, but rather as a smooth
crossover.

\begin{figure}
\includegraphics[clip,height=2.7in]{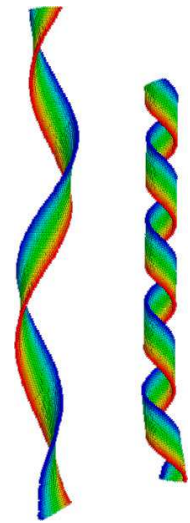}%
\caption{(Color online) Two ribbons grown with the same elastic parameters, but
with different orientations of the director representing the molecular tilt. In
the left case, the tilt is aligned along the long axis of the ribbon, but in the
right case it is aligned along the short axis. This change in the tilt direction
reverses the handedness of the ribbon, and greatly changes the radius and
pitch.}
\end{figure}

To further investigate this model, we explore the consequences of changing the
molecular tilt direction in the membrane. In the original simulations, the
director was constrained to point along the ribbon's long edge, along a bond
between neighboring interaction sites (as in a smectic-$I$ liquid crystal). We
now rotate it by $90^\circ$ to point perpendicular to the long edge, halfway
between two nearest-neighbor bonds (as in a smectic-$F$ liquid crystal).
Although the sign and magnitude of the membrane's chiral interaction parameter
$\lambda$ remain unchanged, the handedness of the aggregate switches from right-
to left-handed, as shown in Fig.~4. The pitch and diameter of the aggregate also
change greatly.

While somewhat counter-intuitive, the handedness reversal is physically
reasonable. Earlier theoretical work showed that the curvature direction favored
by molecular chirality is $45^\circ$ from the tilt direction~\cite{selinger01},
so rotating the tilt direction by $90^\circ$ should change the curvature
direction by $90^\circ$, giving a handedness reversal. (As a thought experiment,
we can imagine cutting a cylindrical tubule to form a helical ribbon. If we cut
it along the tilt direction, the ribbon has one handedness. If we cut it
perpendicular to the tilt direction, the ribbon has the opposite handedness.)
This result shows that aggregates may form with either handedness even from
enantiomerically pure material, simply by variation of the molecular tilt
direction with respect to the edges. This could be a possible explanation of the
experiments on conflicting handedness mentioned earlier~\cite{thomas99}.

The change in pitch and diameter is a more subtle effect, which we cannot
attribute to the change in tilt direction relative to the ribbon's long edge.
Rather, it is probably due to changing the tilt direction relative to the
nearest-neighbor bonds in the membrane. This change can greatly modify the
elastic properties of the membrane, which would alter the geometric parameters
of the resulting aggregate. This result shows that aggregates can have different
pitch and diameter even if they are composed of the same material, just by
variation of the tilt direction with respect to the crystallographic axes.  This
could be a possible explanation of the different geometric structures seen in
bile~\cite{chung93,zastavker99}.

In conclusion, we have developed an accelerated approach for studying the
shapes of chiral aggregates through Monte Carlo simulation of a growth process.
The simulation results show that a short-range chiral interaction leads to the
formation of complex chiral shapes, which undergo a gradual crossover between
Gaussian saddle-like curvature and cylindrical curvature as a function of the
interaction parameters. The handedness of the macroscopic shapes is determined
by the interplay between the microscopic chiral energetic parameter and the
orientation of the molecular tilt. Thus, this work shows the range of shapes
that occur in experimental studies of chiral self-assembly, provides a possible
explanation for experiments on conflicting handedness, and demonstrates an
approach for future simulations based on even more detailed microscopic models
of chiral molecular structure.

This work was supported by the Office of Naval Research
and the National Science Foundation Grant No.\ DMR-0116090. APM was
supported by a fellowship from the National Research Council.

\end{document}